\documentclass[reqno,a4paper,12pt,final]{amsart}
\usepackage[utf8x]{inputenc}
\usepackage[a4paper,hmargin=2cm,tmargin=3cm,bmargin=3cm]{geometry}
\usepackage{amssymb,amstext,amsthm,eucal}
\usepackage{color}
\usepackage{array}
\usepackage[all]{xy}
\usepackage[nosort]{cite}
\usepackage{hyperref}
\hypersetup{%
  pdftitle   = {On the maximal superalgebras of supersymmetric backgrounds},
  pdfkeywords = {supergravity, superalgebra, maximal superalgebra, M-superalgebra, maximally supersymmetric},
  pdfauthor  = {José Figueroa-O'Farrill, Emily Hackett-Jones, George Moutsopolous, Joan Simón},
  pdfcreator = {\LaTeX\ with package \flqq hyperref\frqq}
}
\PrerenderUnicode{éÉóÓ}
\newcommand{\half}{\tfrac12}
\newcommand{\be}{\boldsymbol{e}}

\newcommand{\fg}{\mathfrak{g}}
\newcommand{\fk}{\mathfrak{k}}
\newcommand{\fm}{\mathfrak{m}}
\newcommand{\fz}{\mathfrak{z}}

\newcommand{\fs}{\mathfrak{s}}

\newcommand{\fso}{\mathfrak{so}}
\newcommand{\fco}{\mathfrak{co}}
\newcommand{\fiso}{\mathfrak{iso}}
\newcommand{\fosp}{\mathfrak{osp}}

\newcommand{\Cl}{\mathrm{C}\ell}

\newcommand{\fspin}{\mathfrak{spin}}

\newcommand{\Spin}{\mathrm{Spin}}
\newcommand{\SO}{\mathrm{SO}}

\ifx\Sp\UnDeFiNeD
\newcommand{\Sp}{\mathrm{Sp}}
\else
\renewcommand{\Sp}{\mathrm{Sp}}
\fi

\newcommand{\RR}{\mathbb{R}}

\DeclareMathOperator{\End}{End}

\DeclareMathOperator{\dvol}{dvol}

\DeclareMathOperator{\AdS}{AdS}

\newcommand{\unity}{\boldsymbol{1}}
\newcommand{\ket}[1]{{#1}}
\newcommand{\braket}[2]{\left({#1},{#2}\right)}
\newcommand{\bra}[1]{\overline{#1}}
\newcommand{\ketbra}[2]{\ket{#1}\,\bra{#2}}
\newcommand{\MUNCH}[1]{\relax}

\allowdisplaybreaks[1]
\begin{document}
\title{On the maximal superalgebras of supersymmetric backgrounds}
\author[Figueroa-O'Farrill]{José Figueroa-O'Farrill}
\author[Hackett-Jones]{Emily Hackett-Jones}
\author[Moutsopoulos]{George Moutsopoulos}
\author[Simón]{Joan Simón}
\address{School of Mathematics and Maxwell Institute for Mathematical Sciences, University of Edinburgh, James Clerk Maxwell Building, King's Buildings,
  Edinburgh EH9 3JZ, UK}
\address[JMF also]{Departament de Física Teòrica \& IFIC (CSIC-UVEG), Universitat de València, 46100 Burjassot, Spain}
\thanks{EMPG-08-15}
\begin{abstract}
  In this note we give a precise definition of the notion of a maximal superalgebra of certain types of supersymmetric supergravity backgrounds,
  including the Freund--Rubin backgrounds, and propose a geometric construction extending the well-known construction of its Killing superalgebra.  We
  determine the structure of maximal Lie superalgebras and show that there is a finite number of isomorphism classes, all related via contractions from
  an orthosymplectic Lie superalgebra.  We use the structure theory to show that maximally supersymmetric waves do not possess such a maximal
  superalgebra, but that the maximally supersymmetric Freund--Rubin backgrounds do.  We perform the explicit geometric construction of the maximal
  superalgebra of $\AdS_4\times S^7$ and find that is isomorphic to $\fosp(1|32)$.  We propose an algebraic construction of the maximal superalgebra of
  any background asymptotic to $\AdS_4 \times S^7$ and we test this proposal by computing the maximal superalgebra of the M2-brane in its two maximally
  supersymmetric limits, finding agreement.
\end{abstract}
\maketitle
\tableofcontents

\section{Introduction}
\label{sec:intro}

It has been known for some time that maximally extended superalgebras encode information about BPS states interpretable as brane and/or wave excitations
above the vacuum annihilated by all of its generators \cite{QuimMbrane}.  Although this claim is expected to hold in general, it has been mostly
realised in asymptotically flat spacetimes (using gravitational terminology) or in superPoincaré invariant field theories in diverse dimensions
\cite{QuimMbrane, TownsendMTfS, Sato1, Sato2, Crapsetal}.  In stringy terms, the vacuum of all these superalgebras can be interpreted either in terms of
the Minkowski vacuum itself or as a configuration of intersecting branes.

There are several reasons why these claims are difficult to extend to general backgrounds.  From a gravity perspective, it is not known in general how
to define the notion of conserved charges for arbitrary asymptotics.  From a field theory perspective, supersymmetry methods are mainly based on
superPoincaré or superconformally invariant theories.

In this work we follow an entirely geometrical approach, extending the one giving rise to the Killing superalgebra --- see, e.g., \cite{FMPHom}.  The
idea is simple: the bosonic generators of the Killing superalgebra are geometrically Killing vectors which can be constructed by squaring Killing
spinors, but they are not the only objects which can be constructed in this way.  In general one can construct also higher-rank differential forms and
it is therefore natural to include them in an extension of the Killing superalgebra, a larger algebraic structure we will call the maximal superalgebra
(see Section \ref{sec:structure} for a precise definition).  The idea itself is not new, appearing already in \cite{GeorgeKillingForms}, which stops
short of the calculation of the superalgebra, while correctly observing that the extra bosonic generators in the superalgebra will fail in general to be
central.  As we will see below, this is to be expected from the algebraic structure of maximal superalgebras.

The main difficulty in this approach lies not in the computation of these extra bosonic generators, which can be determined after all using the
representation theory of the isometry group to decompose the symmetric square of the odd generators, but rather in proving that the resulting extended
algebraic structure is a Lie superalgebra. This is a point that is not perhaps widely appreciated, since part of the literature assumes that one can
define an action of the new bosonic generators on the odd generators such that all Jacobi identities are satisfied.  This is, however, nontrivial.  This
can be traced to the fact that the gravitino connection defining the notion of a Killing spinor is not in general induced from a connection on the
tangent bundle.  As a result, although the connection is defined on the bundle of differential forms, it does not respect the grading.  Therefore the
differential forms obtained by squaring Killing spinors will obey complicated partial differential equations which involve all the forms at once --- the
exception being the one-forms which are dual to the Killing vectors which are the bosonic generators of the Killing superalgebra.  There are at present
no results which tell us how such ``supergravity Killing forms'' act on Killing spinors and hence no way to define some of the brackets in the putative
maximal superalgebra.

One way to overcome this difficulty is to restrict the class of backgrounds on which we work to those for which the gravitino connection does come
induced from a metric affine connection.  One such class is given by the Freund--Rubin backgrounds.  In those backgrounds the fluxes are geometric ---
given by volume forms of lower-dimensional manifolds of which the spacetime is a product --- and the Killing spinors are made out of geometric Killing
spinors on each of the factors.  The simplest Freund--Rubin backgrounds are those which are maximally supersymmetric and in this note we will
concentrate, for simplicity, on $\AdS_4\times S^7$.  Our results certainly extend to the other maximally supersymmetric Freund--Rubin backgrounds and
very likely also to Freund--Rubin backgrounds preserving less supersymmetry.

The primary tool of our analysis is the well-known cone construction \cite{Baer} of Bär's, which establishes a one-to-one correspondence between
geometric Killing spinors on a spin manifold and parallel spinors on its metric cone.  Its use in the construction of the Killing superalgebra was
pioneered in \cite{AFHS} and explained in \cite{JMFKilling}.  Once the Killing superalgebra is lifted to the cone and the action of the Killing vectors
(which lift to parallel 2-forms) understood in terms of Clifford multiplication, it provides us with a natural extension for any differential form
appearing in the symmetric square of supergravity Killing spinors.  Finally, once it is proved that the extended algebraic structure closes into a Lie
superalgebra, we project onto the base of the cone and we learn how the even-odd and even-even brackets had to be geometrically defined, thus extending
the action of Killing vector fields on Killing spinors and the standard Lie algebra structure of Killing vector fields.

This method would perhaps be of marginal interest at best if the only accessible supergravity backgrounds besides the Minkowski vacua were the maximally
supersymmetric Freund--Rubin backgrounds.  We will see that in fact the method extends (algebraically) to define the maximal superalgebras of less
supersymmetric supergravity backgrounds which are asymptotic to them.  We will illustrate this with the case of the M2-brane, an interpolating soliton
connecting the eleven-dimensional Minkowski vacuum and $\AdS_4 \times S^7$.  The Killing spinors of the M2 brane define subspaces of the space of
Killing spinors of the Minkowski vacuum and of $\AdS_4 \times S^7$ and hence they generate a sub-superalgebra of the Lie superalgebras associated with
their Killing spinors.  As a check of the construction, we verify that the sub-superalgebras of the maximal superalgebras of the Minkowski vacuum and
of $\AdS_4 \times S^7$ obtained in this way are indeed isomorphic.

The rest of this note is organised as follows.  In Section~\ref{sec:Killing} we introduce the different superalgebras associated to the Killing spinors
of a supersymmetric supergravity background and illustrate this with the well-known case of the eleven-dimensional Minkowski vacuum.  In
Section~\ref{sec:structure} we describe the structure theory of maximal superalgebras and introduce some notation for different kinds of Lie
superalgebras which will appear in the rest of the paper.  We will see that maximal superalgebras are extremely constrained and consist of a finite
number of isomorphism classes which, as shown in the appendix, can be obtained from an orthosymplectic Lie superalgebra via contractions.  This rigidity
is very useful in order to derive (non)existence results and we illustrate this with the result that the maximally supersymmetric waves do not admit
maximal superalgebras.  In Section~\ref{sec:AdS4S7} we describe in detail the geometric construction of the maximal superalgebra of $\AdS_4 \times S^7$
geometrically after showing using the structural results of the preceding section that, if it exists, it must be isomorphic to $\fosp(1|32)$.  In
section~\ref{sec:m2} we describe the Lie superalgebras associated to the M2-brane background and how they relate to the Lie superalgebras of the
interpolated vacua.  As a check of our construction of the maximal superalgebra of $\AdS_4 \times S^7$, we show that the maximal superalgebra of the
M2-brane constructed as a sub-superalgebra of $\fosp(1|32)$ is isomorphic to the one constructed out of the M-algebra of the Minkowski vacuum.  Finally
in Section~\ref{sec:discuss}, we discuss our results and compare with the existing literature.

\section{Lie superalgebras associated to Killing spinors}
\label{sec:Killing}

The Killing spinors of a supersymmetric supergravity background satisfy linear equations (the vanishing of the variations of the fermionic fields in the
theory) and hence they define a vector space which we denote by $\fk_1$.  This vector space is the odd subspace of several superalgebras naturally
associated to the background.

Firstly we have the \textbf{Killing superalgebra} $\fk = \fk_0 \oplus \fk_1$, where $\fk_0 = [\fk_1,\fk_1]$ consists of those infinitesimal
automorphisms of the bosonic background which can be constructed by squaring spinors.  The elements of $\fk_0$ are Killing vectors leaving invariant all
the other bosonic fields which are turned on in the background.  Any other such Killing vectors not in $\fk_0$ act on the Killing superalgebra as outer
derivations.  Adding them to the Killing superalgebra we arrive at the \textbf{symmetry superalgebra} of the background, of which the Killing
superalgebra is clearly an ideal.  If we let $\fs = \fs_0 \oplus \fs_1$ denote the symmetry superalgebra, then $\fs_1 = \fk_1$ and $\fs_0$ consists of
all Killing vectors leaving invariant the bosonic fields turned on in the background.  Both the Killing and symmetry superalgebras are known to be Lie
superalgebras.  This has been shown for the symmetry superalgebra in \cite{FMPHom,EHJGMHom} for the eleven- and ten-dimensional supergravities,
respectively, but it is true more generally, whereas the result for the Killing superalgebra follows from this one.

For example, the Killing superalgebra of the Minkowski background is the supertranslation ideal of the Poincaré superalgebra.  This is easy to see
because the Killing spinors $Q_i$ are parallel and so are any objects constructed covariantly out of them.  However the symmetry superalgebra is the
Poincaré superalgebra itself, which in addition to the translations $P_\mu$ also contains the Lorentz generators $M_{\mu\nu}$.  We may extend the
Killing superalgebra by the addition of central charges.  The maximal such extension is called the M-algebra in \cite{TownsendMTfS} for supergravity
theories with $32$ supercharges.  We may also extend the M-algebra further by adding the Lorentz generators.  This corresponds to the maximal extension
of the symmetry superalgebra.  The various superalgebras of the Minkowski background in eleven dimensions are shown in table~\ref{tab:minkal}.

\begin{table}
  \centering
  \renewcommand{\arraystretch}{1.5}
  \begin{tabular}{|c|c|c|}\hline
    superalgebras & even basis & bracket $\{Q_i,Q_j\}$ \\\hline\hline
    Killing& $P_\mu$ & $P_\mu (C\gamma^\mu)_{i j}$ \\\hline
    symmetry& $P_\mu$, $M_{\mu\nu}$ & $P_\mu (C\gamma^\mu)_{i j}$ \\\hline
    M-algebra
    & $P_\mu$, $Z_{\mu\nu}$, $\overline{Z}_{\mu\nu\rho\sigma\tau}$
    & $\begin{array}{r}
      P_\mu (C\gamma^\mu)_{i j} + Z_{\mu\nu} (C\gamma^{\mu\nu})_{ij}\\+ \overline{Z}_{\mu\nu\rho\sigma\tau} (C\gamma^{\mu\nu\rho\sigma\tau})_{ij}
    \end{array}$
    \\\hline
    maximal extension& $P_\mu$, $M_{\mu\nu}$, $Z_{\mu\nu}$, $\overline{Z}_{\mu\nu\rho\sigma\tau}$
    & $\begin{array}{r}
      P_\mu (C\gamma^\mu)_{i j} + Z_{\mu\nu} (C\gamma^{\mu\nu})_{ij}\\+ \overline{Z}_{\mu\nu\rho\sigma\tau} (C\gamma^{\mu\nu\rho\sigma\tau})_{ij}
    \end{array}$\\\hline
  \end{tabular}
  \vspace{8pt}
  \caption{Superalgebras of the Minkowski background in d=11.}\label{tab:minkal}
\end{table}

Our point of view, which may or may not coincide with that in \cite{TownsendMTfS}, is that the M-algebra is \emph{not} associated to the theory itself
but only to a particular background, here the Minkowski background.  In other words, for us the `M' in `M-algebra' is not the `M' in `M-Theory', but the
one in `Minkowski'.  This point of view prompts us to ask ourselves the following questions: does there exist an analogue of the M-algebra for other
supersymmetric supergravity backgrounds?  and if so, how do we construct it?

It seems clear that such an algebra should be generated by the Killing spinors, and hence it suggests that the bosonic generators correspond to the
different bispinors we can construct out of them.  In other words, letting $\fm = \fm_0 \oplus \fm_1$ denote this putative superalgebra, we set $\fm_1 =
\fk_1$ and $\fm_0 \cong S^2\fm_1$, with the Lie bracket $S^2\fm_1 \to \fm_0$ defined to be the above isomorphism.  This leaves us with the tasks of
extending the known Lie bracket $\fk_0 \otimes \fm_1 \to \fm_1$ to all of $\fm_0$, the Jacobi identity then fixing the Lie bracket $\Lambda^2\fm_0 \to
\fm_0$, and of checking the remaining components of the Jacobi identity in order to prove that $\fm$ is indeed a Lie superalgebra.  This latter task
being mechanical once the Lie brackets are defined, we concentrate on the former.

In the case of the Minkowski background, $\fk_0$ is central and hence it is consistent to demand that the additional generators in $\fm_0$ be central.
However a cursory glance at the Killing superalgebras of other backgrounds shows this cannot be correct in general.  For example, the Killing
superalgebra of Freund--Rubin backgrounds of the form $\AdS \times S$ contains the full isometry algebra of the background.  If the additional bosonic
generators were central, they would have vanishing brackets with the supercharges and hence also with the isometries, contradicting the fact that they
are tensorial objects.  Therefore we seek a direct geometric construction for the Lie bracket $\fm_0 \otimes \fm_1 \to \fm_1$ which agrees with the
known construction when restricted to $\fk_0$.  In this note we do precisely this for the maximally supersymmetric background $\AdS_4 \times S^7$ of
eleven-dimensional supergravity.  But before embarking on this journey it is convenient to introduce some vocabulary.

\section{Structure of maximal superalgebras}
\label{sec:structure}

In a Lie superalgebra $\fg = \fg_0 \oplus \fg_1$, the component of Lie bracket mapping $S^2\fg_1 \to \fg_0$ need be neither injective nor surjective.
If surjective, $\fg$ is generated by $\fg_1$ and we say that $\fg$ is \textbf{odd-generated}.  For example, the Killing superalgebra of a supersymmetric
supergravity background is (by definition) odd-generated.  In any Lie superalgebra $\fg = \fg_0 \oplus \fg_1$, the odd subspace $\fg_1$ generates an
ideal $[\fg_1,\fg_1] \oplus \fg_1$, which is all of $\fg$ precisely when $\fg$ is odd-generated.  We will say that a Lie superalgebra $\fg = \fg_0
\oplus \fg_1$ is \textbf{full} if the Lie bracket $S^2\fg_1 \to [\fg_1,\fg_1]$ is a vector space isomorphism.  Full algebras need not be odd-generated,
since $\fg_0$ could be strictly larger than its ideal $[\fg_1,\fg_1]$.  A full odd-generated Lie superalgebra will be called \textbf{minimally full} for
the purposes of this note.

The M-algebra is an example of a minimally full superalgebra and indeed, our task can then be described in this language as the construction of a
\textbf{minimally full extension} of the Killing superalgebra of a supersymmetric supergravity background.  We will refer to this putative superalgebra
as the \textbf{maximal superalgebra} of the background, since this agrees morally with the nomenclature used in the literature on this subject.

Minimally full Lie superalgebras have a very simple structure.  Indeed, it follows essentially from \cite[Appendix~A]{KamimuraSakaguchi} that every such
Lie superalgebra $\fg = \fg_0 \oplus \fg_1$ is determined uniquely by a $\fg_0$-invariant skewsymmetric bilinear form $\omega \in \Lambda^2\fg_1^*$.
Relative to a basis $Q_a$ for $\fg_1$ and $Z_{ab} := [Q_a,Q_b]$ for $\fg_0$, the Lie brackets can be written as
\begin{equation}
  \label{eq:minfull}
  \begin{aligned}[m]
    [Z_{ab}, Q_c] &= \omega_{bc} Q_a + \omega_{ac} Q_b\\
    [Z_{ab}, Z_{cd}] &= \omega_{bc} Z_{ad} + \omega_{bd} Z_{ac}
    + \omega_{ac} Z_{bd} + \omega_{ad} Z_{bc}~,
  \end{aligned}
\end{equation}
where $\omega_{ab} = \omega(Q_a,Q_b) = - \omega_{ba}$.  Up to the action of the general linear group, a skewsymmetric bilinear form is uniquely
determined by its rank, which is always an even number: $0,2,\dots,2\lfloor\frac{n}2\rfloor$, where $n = \dim\fg_1$.  Therefore up to isomorphism there
are precisely $\lfloor\frac{n}2\rfloor + 1$ minimally full Lie superalgebras with $\dim\fg_1 = n$.  The most degenerate case is when $\omega = 0$ and
hence the even generators are central.  This is the case for the M-algebra.  At the other extreme and assuming that $n$ is even, we have the case of
nondegenerate $\omega$, which is the orthosymplectic Lie superalgebra $\fosp(1|n)$.  As shown in Appendix~\ref{sec:contractions}, the different
minimally full Lie superalgebras can be obtained from the orthosymplectic one by contractions.  We will show in this note that for the maximal
superalgebra of the maximally supersymmetric Freund--Rubin backgrounds, $\omega$ is symplectic, whence the maximal superalgebra is orthosymplectic.

Assuming the existence of the maximal superalgebra, it is often possible to determine it uniquely (up to isomorphism) using the above results on the
structure of minimally full Lie superalgebras and the fact that it is an extension of the Killing superalgebra.  Indeed, being an extension means that
$\fk_0 < \fm_0$ is a Lie subalgebra and the action of $\fm_0$ on $\fm_1$ restricts to the action of $\fk_0$.  This implies that $\omega$ is, in
particular, $\fk_0$-invariant.  Using representation theory we may determine the space of $\fk_0$-invariants in $\Lambda^2\fm_1^*$, which in many cases,
particularly when $\fk_0$ is large enough, turns out to be one-dimensional, with generator $\omega_0$, say.  This means that a priori and up to
isomorphism there are two minimally full extensions of the Killing superalgebra: $\omega = \omega_0$ and $\omega = 0$.  The latter case corresponds to
$\fm_0$ being central, but that would imply $\fk_0$ to be central as well, contradicting the one-dimensionality of
$\left(\Lambda^2\fm_1^*\right)^{\fk_0}$ (for $\dim\fm_1>2$).  In any case, if $\fk_0$ is not abelian but $\left(\Lambda^2\fm_1^*\right)^{\fk_0}$ is
one-dimensional, then there is a unique minimally full extension of the Killing superalgebra, up to isomorphism.  This argument applies to the maximally
supersymmetric Freund--Rubin backgrounds of eleven-dimensional and IIB supergravities and we are able to conclude that any minimally full extension is
isomorphic to $\fosp(1|32)$.  Of course, even if we find that different backgrounds have isomorphic maximal superalgebras, it is often convenient to
write the maximal superalgebra in a basis in which the covariance under $\fk_0$ is kept manifest.  For the case of $\fosp(1|32)$ this has been done in
\cite{VHVP}.

The construction of the maximal superalgebra of a supersymmetric supergravity background is a special instance of the more general algebraic problem of
constructing the minimally full extension of an odd-generated Lie superalgebra.  We would like to address this algebraic problem here and ask the
question of whether this minimally full extension exists.  Let $\fk = \fk_0 \oplus \fk_1$ be an odd-generated Lie superalgebra, so that $\fk_0 =
[\fk_1,\fk_1]$.  If it exists, a minimally full extension of $\fk$ is Lie superalgebra $\fm = \fm_0 \oplus \fm_1$ with $\fm_1 = \fk_1$, $\fm_0 =
[\fm_1,\fm_1] \cong S^2\fm_1$, $\fk_0 < \fm_0$ is a Lie subalgebra and the restriction of the action of $\fm_0$ on $\fm_1$ to $\fk_0$ coincides with the
action of $\fk_0$ on $\fk_1$ in $\fk$.

Let $(Q_a)$ be a basis for $\fk_1$ and $(K_m)$ be a basis for $\fk_0$, extended to a basis $(K_m,Z_i)$ for $\fm_0$.  We will let $\fz$ denote the span
of $(Z_i)$ so that $\fm_0 = \fk_0 \oplus \fz$.  Since $\fm$ is minimally full, it is given by equation~\eqref{eq:minfull} for some $\omega$ relative to
the basis $Z_{ab} = [Q_a,Q_b]$.  We may write $K_m = \half \phi_m^{ab} Z_{ab}$ in terms of this basis, and then
\begin{equation}
  [K_m,Q_c] = \phi_m^{ab}\omega_{bc} Q_a
\end{equation}
is the action of $\fk_0$ on $\fk_1$.  Let
\begin{equation}
  \fk_1^\perp = \left\{ X \in \fk_1 \middle | \omega(X,Y)=0~\forall Y \in \fk_1 \right\}
\end{equation}
denote the radical of $\omega$.  Then if $Q\in\fk_1^\perp$ then $[K_m,Q]=0$, whence $\fk_1^\perp \subset \fk_1^{\fk_0}$.  If no nonzero element of
$\fk_1$ is invariant under all of $\fk_0$, so that $\fk_1^{\fk_0} = 0$, then $\fk_1^\perp =0$, whence $\omega$ must be symplectic.  At the same time, if
$\omega$ is symplectic, then no nonzero element of $\fk_0$ can act trivially on $\fk_1$.  Therefore if an odd-generated Lie superalgebra
$\fk=\fk_0\oplus \fk_1$ is such that there exists an element $0 \neq Z \in \fk_0$ such that $[Z,Q] = 0$ for all $Q \in \fk_1$ and there does not exist
$0 \neq Q \in \fk_1$ which commutes with all $\fk_0$, then there is no minimally full extension.

This is indeed the case for the Killing superalgebras \cite{FOPflux,NewIIB} of the Kowalski-Glikman \cite{KG} and BFHP \cite{NewIIB} waves.  It follows
from the expression for those algebras that $P_-$ acts diagonally with nonzero eigenvalues, whence $\fk_1^{\fk_0} = 0$, whereas at the same time $P_+$
is central.  Therefore the maximally supersymmetric waves of eleven-dimensional and type IIB supergravities have no maximal superalgebra according to
our definition.

This result may seem surprising because it is a fact that there exist supersymmetric D-branes in these backgrounds \cite{SkenTayl} and there ought to be
a superalgebra which governs the existence of the corresponding BPS states.  Of course, the superalgebra need not be minimally full in the sense
described here.  This lack of ``fullness'' may well be related to the particular causal boundary structure of these backgrounds \cite{BereNast,
  MaroRoss}, which does not allow one to realise all the conserved charges that dimension alone might allow.  It would be interesting to investigate
this phenomenon further by an explicit construction of the supergravity conserved charges extending the formalism developed in \cite{LeWiRoss}.

\section{The maximal superalgebra of $\AdS_4 \times S^7$}
\label{sec:AdS4S7}

In this section we will show that the Killing superalgebra $\fosp(8|4)$ of the Freund--Rubin background $\AdS_4 \times S^7$ admits a minimally full
extension isomorphic to $\fosp(1|32)$.  To do so, we first review the Killing superalgebra construction, both using the geometry of $\AdS_4\times S^7$
and the geometry of the flat cones in $\RR^{2,3}$ and $\RR^8$ where both manifolds can be embedded as quadrics.  Once the Killing superalgebra is
understood in the cone, it provides us with a natural extension for the minimally full superalgebra that keeps all the bosonic generators appearing in
the symmetric square of the Killing spinors.  We prove that the corresponding algebra closes and is isomorphic to $\fosp(1|32)$.  Finally, projecting
the algebraic construction on the cones to $\AdS_4\times S^7$, we learn how to properly define the action of forms on geometric Killing spinors, yielding
the even-odd brackets of the superalgebra and thus extending the existent notion for Killing superalgebras.

\subsection{Background and Killing spinors}

The $\AdS_4\times S^7$ eleven dimensional supergravity background has metric
\begin{equation*}
  ds^2=ds^2(\AdS_4; 8\rho)-ds^2(S^7;7\rho)~,
\end{equation*}
where $8 \rho$ and $7\rho$ are the scalar curvatures of each factor, and the four form field strength supporting it is
$F_{(4)}=\sqrt{6\rho}\dvol(\AdS_4)$.  Its Killing spinors can be expressed in terms of geometric Killing spinors $\varepsilon$ belonging to
the kernel of $(\nabla_X-\lambda X\cdot)$, where the Killing number $\lambda$ of a metric manifold with constant scalar curvature $R$ and dimension $d$
satisfies $R=4\lambda^2 d(d-1)$.  The vector field $X$ acts on the spinor $\varepsilon$ via Clifford multiplication $X\cdot\varepsilon$.  In components,
we have
\begin{equation*}
  \nabla_\alpha \varepsilon = \lambda \Gamma_\alpha \varepsilon\,.
\end{equation*}
A suitable representation of the gamma matrices of $\Cl(1,10)$ is given by
\begin{align*}
  \Gamma^\mu & =\dvol(1,3)\gamma^{\mu}\otimes\unity~,\quad\mu=0,1,\ldots 3\\
  \Gamma^i   & =\dvol(1,3)\otimes \gamma^i~,\quad i=4,5,\ldots 10~, 
\end{align*}
where $\gamma^\mu$ and $\gamma^i$ generate representations of $\Cl(1,3)$ and $\Cl(7)$ respectively.  The Killing spinor equation is then solved by the
tensor product of geometric Killing spinors on each factor: $\varepsilon = \varepsilon_{\text{AdS}}\otimes\varepsilon_{\text{S}}$.  This fixes the odd
part of the Killing superalgebra $\fk_1$.  The even part is the full isometry algebra of the background, $\fk_0=\fso(2,3)\oplus \fso(8)$, and is
generated by squaring the Killing spinors to Killing vectors.This defines the odd-odd bracket of the superalgebra.  The even-odd bracket of the Killing
superalgebra is defined in terms of the action of the Killing vector field $\xi\in \fk_0$ on the Killing spinor $\varepsilon\in\fk_1$
\begin{equation}
  \label{eq:killaction}
  [\xi,\,\varepsilon] \equiv L_\xi\varepsilon = \frac{1}{2} \xi^m\nabla_m\varepsilon - \frac{1}{4}\left(\nabla \xi\right)_{m_1m_2} \Gamma^{m_1m_2}
  \varepsilon~.
\end{equation}
Finally, the even-even bracket is just given by the standard commutator of vector fields.  It can be shown that all Jacobi identities are satisfied given
these definitions \cite{EHJGMHom}.  Thus the Killing superalgebra is a Lie superalgebra.

\subsection{Killing superalgebra from the cone perspective}

Both $\AdS_4$ and $S^7$ are manifolds $M$ with metric $g$ and constant curvature $R$ allowing a description as the base of a flat cone of the manifold
$M\times \mathbb{R}^+$ with metric
\begin{equation*}
  \hat{g}_{\text{cone}} = dr^2 + r^2\,g\,.
\end{equation*}
In particular, both manifolds M are described by quadrics $\hat{g}_{mn}\,\hat{Y}^m\hat{Y}^n = \eta\,(2\lambda)^2$ in (an open subset of) $\RR^{2,3}$ and
$\RR^8$, respectively, and $\eta$ depends on the signature of the metric.

The isometry algebra $\fk_0 = \fso(2,3) \oplus \fso(8)$ acts linearly on the cone.  Furthermore, as explained in the present context in
\cite{JMFKilling}, using this cone construction \cite{Baer,KathHabil}, geometric Killing spinors on $\AdS_4$ and $S^7$ are in one-to-one correspondence
with parallel spinors on the corresponding flat cones:
\begin{equation*}
  \varepsilon_{\text{AdS}},\,\varepsilon_{\text{S}} \leftrightarrow \hat{\varepsilon}_{\text{AdS}},\,\hat{\varepsilon}_{\text{S}} \quad \quad
  \hat{\nabla}_{\hat{X}}\,\hat{\varepsilon}_{\text{AdS}} = \hat{\nabla}_{\hat{X}}\,\hat{\varepsilon}_{\text{S}}  = 0\,.
\end{equation*}
As a representation of the isometry algebra $\fk_0 = \fso(2,3) \oplus \fso(8)$, the Killing spinors defining the odd part of the superalgebra $\fk_1$
are $\fk_1 = \Delta_A \otimes \Delta_S^+$ where $\Delta_A$ is the real 4-dimensional spinor irreducible representation of $\Spin(2,3)$ and
$\Delta_S^+$ is the real 8-dimensional positive-chirality spinor irreducible representation of $\Spin(8)$.

On the other hand the dual 1-form $\xi$ of a Killing vector on a quadric like $\AdS_4$ or $S^7$
satisfies a special Killing integrability condition:
\begin{equation}
  \nabla_X d\xi=-8\lambda^2 X^\flat\wedge\xi~,
  \label{eq:killintegral}
\end{equation}
where $X^\flat$ stands for the canonical dual form of the vector field $X$.  Such special Killing 1-forms lift to parallel 2-forms $\hat\xi$ on the
cone; that is, $\hat\nabla\,\hat\xi=0$.  The connection can be made more explicit by using the quadrics $\hat{g}_{mn}\,r^2\,Y^m\,Y^n =
\eta\,(2\lambda)^2$, where we used natural coordinates on the cone $\hat{Y}^m = r\,Y^m$.  Parallel 2-forms are then given in terms of $\hat\xi =
d\hat{Y}^m\wedge d\hat{Y}^n = d\left(r^2 \xi\right)$, for any pair $(m,\,n)$.  In this representation, $2\xi = Y^m\,dY^n - Y^n\,dY^m$.  Thus, the even
elements of the superalgebra $\fk_0$ are described in terms of $\Lambda^2_A\oplus \Lambda^2_S$, the set of 2-forms in $\RR^{2,3}$ and $\RR^8$,
respectively.  Notice the dimensionality of these vector spaces matches the dimension of the corresponding isometry Lie algebras $\fso(2,3) \oplus
\fso(8)$, as they should.

We can now describe the geometric construction of the Killing superalgebra from the perspective of the cone.  The odd-odd bracket is the natural
symmetric square of parallel spinors $\hat\varepsilon$ onto parallel 2-forms $\hat\xi$
\begin{equation*}
  S^2\fk_1\rightarrow \fk_0=\Lambda^2_A \oplus \Lambda^2_S~,
\end{equation*} 
on the respective flat cones.  The even-odd bracket is defined in terms of the action of the 2-forms on the parallel spinors.  This is given by Clifford
multiplication
\begin{equation*}
  [\hat\xi,\,\hat\varepsilon ] \equiv \hat\xi\cdot\hat\varepsilon\,.
\end{equation*}
When expressing this action in terms of the Clifford algebra on the base, it coincides (up to a factor) with the action of the Lie derivative along the
Killing vectors $\xi$ on the Killing spinors $\varepsilon$ given in \eqref{eq:killaction}.  Finally, the even-even bracket is the Clifford commutator on
the cone and agrees with the standard vector field Lie bracket when projecting onto the base.  We are thus able to recast the proof of the closure of the
Killing superalgebra in terms of Clifford multiplication actions on the cone.

\subsection{Maximal superalgebra extension on the cone}

The cone construction allows us to immediately investigate if and how the Killing superalgebra can be fully extended.  This is because while keeping the
identification of all the odd generators, i.e. $\fm_1=\fk_1$, their squared bilinears are not restricted to parallel 2-forms, but include other forms,
as $\fk_0$-representations,
\begin{equation}\label{eq:S2m1FR}
  \begin{aligned}[m]
    S^2\fm_1 &= (S^2\Delta_A \otimes S^2\Delta_S^+) \oplus (\Lambda^2\Delta_A \otimes \Lambda^2\Delta_S^+)\\
    &= \Lambda^2_A \oplus \Lambda^2_S \oplus \left(\Lambda^2_A \otimes \Lambda^{4,+}_S\right) \oplus \left( \Lambda^4_A \otimes \Lambda^2_S\right)~,
  \end{aligned}
\end{equation}
where $\Lambda_A^p$ and $\Lambda_S^p$ are the $p$-th exterior powers of the vector representations of $\fso(2,3)$ and $\fso(8)$, respectively, with the
subscript $+$ denoting the self-dual forms.  The first summands in the last line of the above equation correspond to $\fk_0$ and the other two to
extended bosonic charges.  Perhaps not surprisingly, all bosonic charges constructed in this way turned out to be (the product of) parallel $(p+1)$-forms
$\hat\xi^{(p)}$, where $p=3$ corresponds to the extended bosonic charges both in $\AdS_4$ and the $7$-sphere.

We shall connect these parallel forms to forms on the base shortly, but first, we want to identify what maximal superalgebra this construction is
isomorphic to, in the language of Section~\ref{sec:structure}, and whether it closes a superalgebra at all.  Remember minimally full superalgebras are
characterised by the rank of an skewsymmetric bilinear $\omega$ which is $\fk_0$-invariant.  One computes
\begin{equation}
  \begin{aligned}[m]
    \Lambda^2\fm_1 &= (S^2\Delta_A \otimes \Lambda^2\Delta_S^+) \oplus (\Lambda^2\Delta_A \otimes S^2\Delta_S^+)\\
    &= \RR \oplus \Lambda^1_A \oplus \Lambda_S^{4,+} \oplus \left(\Lambda^1_A \otimes \Lambda_S^{4,+}\right) \oplus \left(\Lambda^2_A \otimes
      \Lambda^2_S\right)~,
  \end{aligned}
\end{equation}
which shows that $\left(\Lambda^2\fm_1^*\right)^{\fk_0}$ is one-dimensional.  (We use that $\fm_1 \cong \fm_1^*$ as $\fk_0$-representations.)  Since
$\fm_1^{\fk_0} = 0$, $\omega$ is symplectic.  In fact, $\omega$ is none other than the Spin-invariant inner product on spinors.  Therefore any minimally
full extension of $\fosp(8|4)$ must be isomorphic to $\fosp(1|32)$ and hence if the maximal superalgebra of $\AdS_4 \times S^7$ does exist, it must be
isomorphic to $\fosp(1|32)$.

The way we define both even-odd and even-even brackets is through Clifford action multiplication, extending the action we described for the Killing
superalgebra.

\subsection{Checking Jacobi identities}

To prove our brackets satisfy all the Jacobi identities, we will introduce some notation and some slightly more mathematical abstractions that will
facilitate our task.  We will let $\alpha\otimes\beta, \alpha'\otimes\beta'\in \Delta_A\otimes\Delta^+_S$ be elements in $\fm_1$, whose bracket
\begin{equation*}
  [ \ket{\alpha}\otimes\ket{\beta} , \ket{\alpha'}\otimes\ket{\beta'} ] =
  \ketbra{\alpha}{\alpha'}\otimes\ketbra{\beta}{\beta'}+\ketbra{\alpha'}{\alpha}\otimes\ketbra{\beta'}{\beta}
\end{equation*}
is then an element of $\fm_0$.  It is useful to think of $\ketbra{\alpha}{\alpha'}$ as the rank-one endomorphism of $\Delta_A$ sending, say,
$\alpha''\mapsto \braket{\alpha'}{\alpha''}\ket{\alpha}$ and similarly for $\ketbra{\beta}{\beta'}\in\End(\Delta_S)$.  The action of these endomorphisms
is naturally defined in terms of the spinors inner products $\braket{-}{-}$, which is symplectic on $\Delta_A$ and symmetric on $\Delta_S$.  Notice that
by using Fierz identities, these endomorphisms $\ketbra{\alpha}{\alpha'}$ and $\ketbra{\beta}{\beta'}$ can be explicitly written in terms of spinor
bilinears involving conveniently antisymmetrised products of Clifford matrices, matching our previous bosonic charge discussion.

The even-odd bracket is the action of $\fm_0$ on $\fm_1$
\begin{equation*}
  [ [ \ket{\alpha}\otimes\ket{\beta} , \ket{\alpha'}\otimes\ket{\beta'} ], \ket{\alpha''}\otimes\ket{\beta''} ] = \braket{\alpha'}{\alpha''}
  \braket{\beta'}{\beta''} \ket{\alpha} \otimes \ket{\beta} + \braket{\alpha}{\alpha''} \braket{\beta}{\beta''} \ket{\alpha'}\otimes\ket{\beta'}~,
\end{equation*}
and the even-even bracket is the commutator of two such endomorphisms.  In the current discussion, we see the symplectic form $\omega$ discussed above
is none other that the inner product on $\Delta_A\otimes\Delta_S^+$.  One can show immediately that the endomorphisms in $\fm_0$ are skew with respect
to $\omega$, which define the even part of $\fosp(1|32)$.  Let us now verify that the algebra closes under the above brackets.

The odd-odd-odd Jacobi identity is satisfied due to the antisymmetry properties of the inner product.  We can polarise this identity for a single odd
element $\ket{\alpha}\otimes\ket{\beta}$, for which
\begin{equation*}
  [[\ket{\alpha}\otimes\ket{\beta},\ket{\alpha}\otimes\ket{\beta}],\ket{\alpha}\otimes\ket{\beta}]=0~.
\end{equation*}

The even-odd-odd Jacobi identity can be shown, again by polarising for a single odd element $\alpha\otimes\beta$ and for an even element
$[\ket{\alpha'}\otimes\ket{\beta'},\ket{\alpha''}\otimes\ket{\beta''}]$.  We then expand
\begin{equation*}
  \begin{aligned}[2]
    [ [\ket{\alpha'}\otimes\ket{\beta'},\ket{\alpha''}\otimes\ket{\beta''}], [ \ket{\alpha}\otimes\ket{\beta}, \ket{\alpha}\otimes\ket{\beta} ]]&=2
    \braket{\alpha''}{\alpha}\braket{\beta''}{\beta}\ketbra{\alpha'}{\alpha}\otimes\ketbra{\beta'}{\beta} \\
    &\quad  + 2 \braket{\alpha'}{\alpha}\braket{\beta'}{\beta}\ketbra{\alpha''}{\alpha}\otimes\ketbra{\beta''}{\beta}\\
    &\quad - 2 \braket{\alpha}{\alpha'}\braket{\beta}{\beta'}\ketbra{\alpha}{\alpha''}\otimes\ketbra{\beta}{\beta''}\\
    &\quad  - 2 \braket{\alpha}{\alpha''}\braket{\beta}{\beta''}\ketbra{\alpha}{\alpha'}\otimes\ketbra{\beta}{\beta'}\\
    & = 2[ \ket{\alpha}\otimes\ket{\beta} , [[\ket{\alpha'}\otimes\ket{\beta'},\ket{\alpha''}\otimes\ket{\beta''}],\ket{\alpha}\otimes\ket{\beta}]]~.
  \end{aligned}
\end{equation*}
Finally the even-even-odd and even-even-even Jacobi identities follow from our construction of $\fm_0$ as (skew-)endomorphisms of $\fm_1$.

Having checked the closure of our algebra, we conclude it is a Lie superalgebra, isomorphic to $\fosp(1|32)$, corresponding to the minimal full
extension of the Killing superalgebra $\fosp(8|4)$.

\subsection{Maximal superalgebra from the $\AdS_4\times S^7$ perspective}

When we reviewed the Killing superalgebra for $\AdS_4\times S^7$, we lifted its Killing vectors to parallel 2-forms on the respective flat cones.  Once
we have the minimally full superalgebra, including the extended bosonic charges, it is natural to provide a geometrical description of them in terms of
the base of the cone, that is, in terms of the geometry of $\AdS_4$ and $S^7$.

All forms generated from the symmetric squaring of parallel Killing spinors on the cone give rise to parallel even forms.  Any such $(p+1)$-form
$\hat{\xi}^{(p)}$ satisfies the condition $\hat{\xi}^{(p)} = d\left(r^{p+1}\,\xi^{(p)}\right)$, which identifies the corresponding odd form on the base
as $\xi^{(p)}$.  These forms satisfy the following two equations, coming from its parallel character on the cone, with $p=2n+1$:
\begin{equation*}
  \begin{aligned}[m]
    \nabla_X \xi^{(2n+1)}&=-2\lambda\iota_X \xi^{(2n+2)}~,\\
    \nabla_X \xi^{(2n+2)}&=2\lambda X^\flat\wedge\xi^{(2n+1)}~.
  \end{aligned}
\end{equation*}
The first equation generalises the notion of (the dual of) a Killing vector to higher-degree forms.  However, the second equation is a special form of
Killing's integrability condition, as in \eqref{eq:killintegral}.  We call such Killing forms \textbf{special}.  In spaces of constant curvature like
$\AdS_4$ or $S^7$ all Killing forms are special.  (\emph{Proof}: Killing's generalised integrability condition bounds the dimension of the space of
Killing $p$-forms on a manifold of dimension $d$ above by $\binom{d+1}{p+1}$, which is precisely the dimension of the space of parallel forms on a flat
cone which descend to special Killing forms.)  These statements are equivalent to saying that the symmetric square of two geometric Killing spinors
$\varepsilon$ consists of a special Killing forms $\xi^{(2n+1)}$ of odd degree, whose covariant derivative is proportional to the even-degree forms
$\xi^{(2n+2)}$.  Finally, the scalars produced from Killing spinors are constants in the base.

If we denote $\Gamma^n_A$ the space of special Killing $n$-forms over $\AdS_4$, the symmetric square of Killing spinors spans $\Gamma^1_A$ and the
exterior square spans $\mathbb{R}\oplus\Gamma^3_A$.  Likewise for the 7-sphere, the symmetric square spans $\mathbb{R}\oplus\Gamma^{3+}_S$, where
$\Gamma^{3+}_S$ contains those special Killing three-forms whose covariant derivative is proportional to their Hodge dual, while their exterior square
spans $\Gamma^1_S$.  This way, one recovers the bosonic elements of $\fosp(1|32)$ in equation \eqref{eq:S2m1FR} from the following tensorial objects
\begin{equation}\label{eq:tenskill}
  \Gamma^1_A\otimes\left(\mathbb{R}\oplus\Gamma^{3+}_S\right)\oplus \left(\mathbb{R}\oplus\Gamma^3_A\right)\otimes\Gamma^1_S~.
\end{equation}

We can check the maximality of the construction by summing the dimensions of the different spaces: $\dim\Gamma^1_A = \dim \fso(2,3)=10$, $\dim\Gamma^1_S
= \dim \fso(8) = 28$, $\dim \Gamma^3_A=5$ (dimension of space of parallel $4$-forms on the flat cone in $\RR^{2,3}$) and $\dim \Gamma^{3+}_S = 35$
(dimension of space of self-dual parallel $4$-forms on the flat cone in $\RR^8$) to obtain
\begin{align*}
  \dim\Gamma^1_A\left(1+\dim \Gamma^{3+}_S\right) + \left(1+\dim \Gamma^3_A\right)\,\dim\Gamma^1_S  &=10\times (1+35) + (1+5)\times 28\\
  &= 528= \dim S^2\fm_1~.
\end{align*}

Having defined the even-odd bracket on the cone, we can learn what the action of these special Killing $p$-forms is on geometric Killing spinors from
the Clifford action of parallel $(p+1)$-forms on parallel spinors on the corresponding cone.  For special odd-degree Killing forms this generalises the
Lie derivative along Killing vectors \eqref{eq:killaction}: if $\xi$ is an odd-degree Killing $p$-form and $\varepsilon$ a geometric Killing spinor,
then
\begin{equation}\begin{aligned}
    L_{\xi}\varepsilon & :=\frac{1}{(p+1)!}\xi_{m_1 m_2 \cdots m_{p-1}}{}^{m_p} \gamma^{m_1 m_2 \cdots m_{p-1}} \nabla_{m_p} \varepsilon \\
    & -\frac{1}{2} \frac{1}{(p+1)!}(\nabla\xi)_{m_1 m_2\ldots m_{p+1}}\gamma^{m_1 m_2\ldots m_{p+1}}\varepsilon
  \end{aligned}
\end{equation}
is also a geometric Killing spinor.  The even-odd Lie bracket of an element $\xi\otimes\zeta$ in $\Gamma^1_A\otimes\Gamma^{3+}_S\oplus
\Gamma^3_A\otimes\Gamma^1_S$ with a supergravity Killing spinor $\varepsilon_{\AdS}\otimes\varepsilon_{S}$ is the action $L_{\xi}\varepsilon_{\AdS}
\otimes L_{\zeta}\varepsilon_{S}$.

Similarly one can define an algebra of special Killing forms from the Clifford algebra of parallel forms on the cone, through which we defined the
even-even brackets.  This will give a sum of mixed-degree forms on the base.  We observe that for two special Killing forms of degree $p$ and $q$ the
truncation to the special Killing form of degree $p+q-1$ coincides with the output of the Nijenhuis--Schouten bracket
\begin{equation*}
  [\xi^{(p)},\xi'^{(q)}]_{\text{NS}}:= 
  (-1)^p g^{\mu\nu}\iota_{\mu}\xi^{(p)}\wedge\nabla_\nu \xi'^{(q)}
  -(-1)^{(p-1)q}g^{\mu\nu}\iota_{\mu}\xi'^{(q)}\wedge\nabla_\nu \xi^{(p)}~,
\end{equation*}
thus proving that the Nijenhuis--Schouten bracket closes on special Killing forms defined on manifolds with constant scalar curvature.  This generalises
the theorem in \cite{Kastor:2007tg} that a sufficient condition for the Nijenhuis--Schouten bracket to close on Killing forms is that the manifold is of
constant sectional curvature.  The cone of such a manifold is locally flat and therefore all Killing forms are special.  However this bracket closes on
any constant scalar curvature manifold, provided we restrict to \emph{special} Killing forms.

\subsection{$AdS_4\times S^7$ superalgebras}

In this section we summarise the construction of the maximal extension $\fosp(1|32)$ of the $\AdS_4\times S^7$ Killing superalgebra. The simplest way to
realise this is by using the tensor product of parallel spinors on the cones $\RR^{2,4}\times\RR^{8}$, a basis of which is $\alpha_i\otimes\beta_I$ with
$i=1\ldots 4$ and $I=1,\ldots 8$. Their symmetric square span their endomorphisms that are skew with respect to the symplectic inner product
$\omega=C\otimes\unity$. Equivalently, the even elements are the parallel forms that appear in the decomposition of equation \eqref{eq:S2m1FR} :
\begin{itemize}
  \item[-] $L_{mn}$ and $L_{ab}$ corresponding to parallel two forms in $\RR^{2,4}$ and $\RR^{8}$, respectively 
  \item[-] $Z^{+}_{mnabcd}$ and $Z_{mnopab}$ corresponding to parallel forms belonging to $\Lambda^2_A \otimes \Lambda^{4,+}_S$ and  $\Lambda^4_A
    \otimes \Lambda^2_S$, respectively.
\end{itemize}
In this basis the algebra is defined using (the tensor product of) Clifford multiplication. The Killing superalgebra is the truncation of the maximal
algebra whereby we keep only the parallel two forms. In this case the Killing superalgebra coincides with the symmetry algebra. The two superalgebras
are shown in table~\ref{tab:FRal}.

\begin{table}
  \centering
  \renewcommand{\arraystretch}{1.5}
  \begin{tabular}{|c|c|c|}\hline
    superalgebras & even basis & bracket $\{\alpha_i\otimes\beta_I,\alpha_{j}\otimes\beta_{J}\}$ \\\hline\hline
    Killing = symmetry &  $L_{mn}$, $L_{ab}$ & 
$\half C_{ij}\gamma^{ab}_{IJ} L_{ab} + 
  \half \delta_{IJ} (C\,\gamma^{mn})_{ij}\,L_{mn}$ \\\hline
    maximal extension & $L_{mn}$, $L_{ab}$, $Z^{+}_{mnabcd}$, $Z_{mnopab}$
    & $\begin{array}{r}
      \half C_{ij}\gamma^{ab}_{IJ} L_{ab} + 
  \half \delta_{IJ} (C\,\gamma^{mn})_{ij}\,L_{mn}\\
+ \frac{1}{2\,4!} \gamma^{ab}_{IJ}\,(C \gamma^{mnop})_{ij} Z_{mnopab}\\ +
\frac{1}{2\,4!} \gamma^{abcd}_{IJ}(C \gamma^{mn})_{ij} Z^{+}_{mnabcd}
    \end{array}$\\\hline
  \end{tabular}
  \vspace{8pt}
  \caption{Superalgebras of $\AdS_4\times S^7$.}\label{tab:FRal}
\end{table}

On the manifold $\AdS_4\times S^7$ the odd elements are the tensor product of geometric Killing spinors and the even elements are the tensor product of
the special Killing forms appearing in \eqref{eq:tenskill}. A suitable basis of the algebra is written using its decomposition under
$\fspin(1,3)\times\fspin(7)$. That is, the even elements are of the form $Z_{\mu_1\ldots \mu_p\,i_1\ldots i_q}$ with flat indices $\mu=0\ldots 3$ and
$i=1\ldots 7$ and the possible pairs $(p,q)=(2,0)$, $(1,0)$, $(0,2)$, $(0,1)$, $(2,3)$, $(1,3)$, $(4,2)$, $(3,2)$, $(4,1)$ and $(3,1)$. The algebra in
this basis can be written explicitly by using (the tensor product of) Clifford multiplication on the base, $\Cl(1,3)\times\Cl(7)$, while the Killing
algebra is the truncation to the forms $Z_{\mu_1}$, $Z_{\mu_1\mu_2}$, $Z_{i_1}$ and $Z_{i_1 i_2}$.

\section{The maximal superalgebra of the M2 brane}
\label{sec:m2}

We may use the determination of $\fosp(1|32)$ as the maximal superalgebra of $\AdS_4 \times S^7$ to compute that of any supergravity background
asymptotic to it.  This is because the asymptotic limits of Killing spinors of a supersymmetric supergravity background are naturally Killing spinors of
the asymptotic background and hence they define a subspace of the Killing spinors of $\AdS_4 \times S^7$.  Once this subspace is identified, it is a
just a matter of algebra to determine the sub-superalgebra of $\fosp(1|32)$ it generates.  Short of an independent geometric calculation of the maximal
superalgebra along the lines advocated here, this algebraic proposal is the only way we know to define the maximal superalgebra of the background.
However for the case of an interpolating soliton we can actually test this proposal, because we can choose either of the interpolated vacua in order to
construct the maximal superalgebra and it had better be that the sub-superalgebras are isomorphic.  In this section we illustrate this method with the
M2-brane of eleven-dimensional supergravity.

The M2-brane is a soliton interpolating between the flat Minkowski background and the Freund--Rubin background $\AdS_4 \times S^7$.  More precisely, the
M2-brane solution is a $2$-parameter family of solutions:
\begin{equation*}
  \begin{aligned}[m]
    g &= H^{-2/3} \eta_3 - H^{1/3} \delta_8\\
    F &= \dvol_\eta \wedge dH^{-1}
  \end{aligned}
\end{equation*}
where $\eta$ and $\delta$ stand for the minkowskian and euclidean metrics, respectively, of dimension given by the subscript, and $H$ is a harmonic
function on the transverse euclidean space of the form
\begin{equation*}
  H = \alpha + \frac{\beta}{r^6}~,
\end{equation*}
where $\alpha,\beta$ are real parameters.  Up to diffeomorphisms and the homothety invariance of the equations of motion, it is only the ratio of the
parameters which matters, whence what we have is a \emph{pencil} of solutions.  The case $\alpha = 0$ corresponds to the Freund--Rubin background
$\AdS_4 \times S^7$ with parameter (scalar curvature) related to $\beta$, and the case $\beta=0$ is Minkowski space.  The Killing spinors of the
background take the form $\varepsilon = H^{-1/6} \varepsilon_\infty$, where $\varepsilon_\infty$ is a parallel spinor in the asymptotic Minkowski
spacetime obeying the ``projection'' condition $\dvol_\eta \cdot \varepsilon_\infty = \varepsilon_\infty$.

The relation between the Killing spinors of the M2-brane background and of the Minkowski background imply, as explained in a somewhat different context
in \cite{FigSimBranes}, that the M2-brane Killing superalgebra $\fk = \fk_0 \oplus \fk_1$ is the sub-superalgebra of the Killing superalgebra of the
Minkowski generated by $\fk_1 = \ker (\dvol_\eta - 1)$.  It is easy to show (see, for example, \cite{FMPHom}) that this implies that $\fk_0$ consists of
translations along the worldvolume of the brane.  Similarly, the symmetry superalgebra of the M2-brane is the normaliser of the Killing superalgebra in
the symmetry superalgebra of the Minkowski background, which is the Poincaré superalgebra.  In other words, $\fs = \fs_0 \oplus \fs_1$, with $\fs_1 =
\fk_1$ and $\fs_0 = \fiso(1,2) \oplus \fso(8)$.  We \emph{define} the maximal superalgebra of the M2-brane as the sub-superalgebra of the M-algebra
generated by $\fk_1$.  If we let $V$ denote the $11$-dimensional vector representation of $\SO(1,10)$ and $V = W \oplus W^\perp$ its decomposition under
the subgroup $\SO(1,2) \times \SO(8)$, where $W$ is the $3$-dimensional vector representation of $\SO(1,2)$ and $W^\perp$ is the $8$-dimensional vector
representation of $\SO(8)$, then the symmetric square of $\fk_1$ decomposes as
\begin{equation}
  \label{eq:S2k1M2}
  S^2\fk_1 = W \oplus \Lambda^2W^\perp \oplus \left(W \otimes \Lambda^4_+ W^\perp\right)~,
\end{equation}
which is a minimally full (central) extension of the Killing superalgebra.  The additional charges are a transverse $2$-form $Z_{ij}$ and a tensor
product $Z_{\mu ijkl}^+$ of a $1$-form on the brane worldvolume and a self-dual transverse $4$-form.  The former is interpreted as describing the charge
of either an M2-brane or an eleven dimensional KK-monopole lying entirely in the transverse space; the latter describes the charge of an M5-brane whose
worldvolume has one dimension along the original M2-brane and the remaining four lie in the transverse space \cite{QuimMbrane}.

We should recover the Killing, symmetry and maximal superalgebras of the M2 as sub-superalgebras of the corresponding superalgebras of $\AdS_4 \times
S^7$.  In particular, for the maximal superalgebra this provides a check on the construction.

The Killing and symmetry superalgebras agree in this case, and are both isomorphic to $\fosp(8|4)$, due to its simplicity, which implies that it is its
own canonical ideal.  The bosonic subalgebra of $\fosp(8|4)$ is isomorphic to $\fso(2,3) \oplus \fso(8)$ and the fermionic subspace is the
representation $\Delta_A \otimes \Delta_S^+$.  We find it convenient to interpret $\fso(2,3)$ as the conformal algebra of $\RR^{1,2}$.  In this guise,
the Killing spinors of the M2-brane are the subspace of $\Delta_A \otimes \Delta_S^+$ which is invariant under the translations.  Let us be very
explicit.

We will choose a pseudo-orthonormal basis $\be_a = (\be_0,\be_1,\be_2,\be_3,\be_4)$ for $\RR^{2,3}$ with $\be_0$ and $\be_4$ timelike.  We will let
$\be_\mu = (\be_0,\be_1,\be_2)$.  Let $L_{ab}$ denote the canonical basis for $\fso(2,3)$.  Then $L_{\mu\nu}$ and $D = L_{34}$ span a $\fco(1,2)$
subalgebra of the conformal algebra of $\RR^{1,2}$.  The translations and the special conformal transformations are given by $P_\mu = L_{3\mu} +
L_{4\mu}$ and $K_\mu = - L_{3\mu} + L_{4\mu}$, respectively.  Introduce null vectors $\be_\pm = \pm \be_3 + \be_4$, whence $P_\mu = L_{+\mu}$ and $K_\mu
= L_{-\mu}$.  The Killing superalgebra of the M2-brane is generated by those spinors which are annihilated by $P_\mu = L_{+\mu}$ or, equivalently, in
the kernel of $\gamma_+ \in \Cl(2,3)$.

Let $\varepsilon_I \otimes \psi_I \in \Delta_A \otimes \Delta_S^+$ for $I=1,2$.  Their bracket in the Killing superalgebra of $\AdS_4 \times S^7$ is
given by
\begin{equation}
  \label{eq:KSAFR}
  [\varepsilon_1 \otimes \psi_1, \varepsilon_2 \otimes \psi_2] = \half (\varepsilon_1, \varepsilon_2)  (\psi_1, \gamma^{ij}\psi_2) L_{ij} + 
  \half (\psi_1, \psi_2) (\varepsilon_1, \gamma^{ab} \varepsilon_2) L_{ab}~,
\end{equation}
where $(-,-)$ stands both for the spin-invariant symplectic form on $\Delta_A$ and the spin-invariant euclidean inner product on $\Delta_S$.  The
Clifford algebra generators $\gamma^a$ for $\Cl(2,3)$ are \emph{symmetric} relative to $(-,-)$, whereas the generators $\gamma^i$ for $\Cl(8)$ are
\emph{skewsymmetric}.  Let $\gamma_+ \varepsilon_I = 0$ for $I=1,2$.  Then in particular, $\varepsilon_1 = \gamma_+ \varepsilon'$ for some $\varepsilon'
\in \Delta_A$.  Hence
\begin{equation*}
  (\varepsilon_1 ,\varepsilon_2) =   (\gamma_+ \varepsilon', \varepsilon_2) =  (\varepsilon' , \gamma_+ \varepsilon_2) = 0~.
\end{equation*}
In other words, the Killing spinors of the M2-brane span a lagrangian subspace of the Killing spinors of the Freund--Rubin background; although they
span a symplectic subspace of the Killing spinors of the Minkowski background, as observed in \cite{FMPHom}.  This is not a contradiction because we are
talking about two different symplectic structures.  A similar argument shows that in $(\varepsilon_1, \gamma^{ab} \varepsilon_2)$ only $(\varepsilon_1,
\gamma^{+\mu} \varepsilon_2)$ is different from zero, hence in \eqref{eq:KSAFR}, only $P_\mu$ appears on the right-hand side.  This shows that the
Killing superalgebra of the M2-brane is a sub-superalgebra of the Killing superalgebra of its near-horizon limit.

The same is true of the symmetry superalgebra: since the symmetry superalgebra is obtained by adding the $\fso(8)$ subalgebra spanned by $L_{ij}$ and
the $\fso(1,2)$ subalgebra spanned by $L_{\mu\nu}$ to the Killing superalgebra.  These generators are also in the symmetry (=Killing) superalgebra of
the Freund--Rubin background and preserve the kernel of $\gamma_+$.  The dilatation $D$ also preserves the kernel of $\gamma_+$, whence the normaliser
in the Freund--Rubin symmetry superalgebra is strictly larger.

Finally, the maximal superalgebra of the M2-brane, defined above, is a sub-superalgebra of the maximal superalgebra of the Freund--Rubin background.
Indeed, we saw above that the M2 Killing spinors span a lagrangian subspace, hence by \eqref{eq:minfull}, the bosonic generators are central.  Under the
subalgebra $\fso(1,2) \oplus \fso(8)$ of $\fosp(1|32)$ the M2 Killing spinors transform according to $\Delta_M \otimes \Delta_S^+$, where $\Delta_M$ is
the two-dimensional irreducible real spinor representation of $\fso(1,2)$ and the bosonic subalgebra of the maximal superalgebra transforms under the
symmetric square, which is again given by \eqref{eq:S2k1M2}.

In summary, the Killing, symmetry and maximal superalgebras of the M2-brane are sub-superalgebras of the Killing, symmetry and maximal (respectively)
superalgebras of its two geometric limits: the asymptotic limit given by Minkowski space and its near-horizon limit given by $\AdS_4 \times S^7$.

\section{Discussion}
\label{sec:discuss}

To summarise, we have shown that the maximal superalgebra of the maximally supersymmetric $\AdS_4\times S^7$ background is isomorphic to $\fosp(1|32)$,
generated by the Killing spinors of the background via a geometric construction which extends the construction of the Killing superalgebra.  At the same
time we have shown that the maximally supersymmetric plane waves do not admit a maximal superalgebra.

There are a number of papers dealing with the maximal superalgebra of curved backgrounds, some of which we agree with, while we seem to be in
disagreement with some others.  First of all we are studying the superalgebra of a \emph{supergravity} background.  Even when some backgrounds,
especially the maximally supersymmetric ones, can be considered as exact string backgrounds, we are not studying the stringy superalgebra, which is
essentially the operator superalgebra of the associated superconformal field theory.  We are therefore not considering fermionic generators other than
than the ones corresponding to the supergravity Killing spinors.  Some papers where extra fermionic charges are considered are
\cite{Peeters:2003vz,AliAkbari:2005is}, but we are unable to compare with their results.  Neither will we compare our results with \cite{Lee:2004jx},
since the algebra in that paper is not maximal.  The proposal for $\fosp(1|32)$ as the the maximal superalgebra of the maximally supersymmetric
Freund--Rubin backgrounds has been made in \cite{Gunaydin:1998km, Gunaydin:1998bt, FerraraPorratiOsp132, Bergshoeff:2000qu, Bergshoeff:2000vg,
  Bandos:2008um}.  In particular, in \cite{FerraraPorratiOsp132} there is a derivation of this fact departing from the algebraic requirement that the
superalgebra be minimally full (in the language of this paper) and that it should contain the maximal superalgebra of the Poincaré superalgebra in the
boundary of $\AdS$.  Several other papers have raised some doubts about this proposal.  For example, in \cite{KamimuraSakaguchi}, which contains the
calculation which makes possible the algebraic characterisation of minimally full Lie superalgebras, it is claimed that $\fosp(1|32)$ is not an
extension of the Killing superalgebra.  A similar claim is made in \cite{MPZAdS}, who then go on to claim that the maximal superalgebra has extra
bosonic generators.  A close inspection of the proposed superalgebra in equation (2.5) in that paper reveals that $\dim[\fg_1,\fg_1] > \dim S^2\fg_1$,
which is impossible as $[\fg_1,\fg_1]$ is the image of a linear map $S^2\fg_1 \to \fg_0$, unless the additional generators are in fact linearly
dependent on the original ones.  The claim in both papers that $\fosp(1|32)$ does not extend the Killing superalgebra is based on an explicit
calculation of the Lie brackets in a choice of basis.

In contrast, in this paper we have \emph{constructed} $\fosp(1|32)$ as an extension of the Killing superalgebra.  The Killing superalgebra is what
remains when we just consider the $1$-forms constructed out of Killing spinor bilinears, whereas $\fosp(1|32)$ is what we obtain when we do no such
projection.  Having shown that the maximally superalgebra is indeed a Lie algebra, we can appeal to the characterisation of such Lie algebras to
identify it with $\fosp(1|32)$ \emph{a posteriori}.  Our methods can be extended to other Freund--Rubin backgrounds, such as $\AdS_7\times S^4$ in
eleven dimensional supergravity or $\AdS_5\times S^5$ in type IIB.

Our initial question whether the Killing superalgebra of a supersymmetric supergravity background admits a minimally full extension seems to have a
negative answer, as suggested by our nonexistence result for the plane waves.  However it may be the case that under some further conditions, e.g.,
existence of a timelike Killing vector, the maximal superalgebra does exist.  It would be interesting to attempt a general geometric construction in
those cases.

\section*{Acknowledgments}

JMF would like to extend his gratitude to José A.~de Azcárraga and to the Departament de Física Teòrica of the Universitat de València for hospitality
and support under the research grant FIS2008-01980 during the final stages of this work.

\appendix

\section{Contractions}
\label{sec:contractions}

In this appendix we will show that all minimally full Lie superalgebras can arise as contractions of the orthosymplectic superalgebra.

Let us recall the notion of a Lie superalgebra contraction.  Let $(\fg= \fg_0 \oplus \fg_1,[-,-])$ be a Lie superalgebra: it is important to distinguish
the underlying vector space $\fg$ from the Lie superalgebra, which is the vector space together with the Lie bracket.  The contraction of a Lie
superalgebra will share the same underlying vector space, but the Lie bracket will be modified.  Let us define what we mean by a contraction of
$(\fg,[-,-])$.  Suppose that $\phi_\varepsilon: \fg \to \fg$ is a family of even linear maps depending on a real parameter $\varepsilon$ with the
property that for all $\varepsilon \neq 0$, $\phi_\varepsilon$ is invertible.  This family of maps defines a new Lie bracket $[-,-]_\varepsilon$ on the
vector space $\fg$ by the requirement that (for $\varepsilon \neq 0$) the map
\begin{equation*}
  \phi_\varepsilon : (\fg, [-,-]) \to (\fg, [-,-]_\varepsilon)
\end{equation*}
be a Lie algebra homomorphism.  Explicitly, for any $X,Y\in\fg$, $[X,Y]_\varepsilon$ is defined by
\begin{equation*}
  [X,Y]_\varepsilon = \phi_\varepsilon^{-1} [\phi_\varepsilon(X), \phi_\varepsilon(Y)]~.
\end{equation*}
By its very definition, the new Lie algebra $(\fg, [-,-]_\varepsilon)$ is isomorphic to the original $(\fg, [-,-])$: the isomorphism being
$\phi_\varepsilon$.  However \emph{suppose} that the limit
\begin{equation*}
  [X,Y]_0 := \lim_{\varepsilon\to 0} [X,Y]_\varepsilon
\end{equation*}
exists for all $X,Y \in \fg$.  By continuity, the skewsymmetry and the Jacobi identity satisfied by $[-,-]_\varepsilon$ for $\varepsilon \neq 0$, will
imply the analogous equations for $[-,-]_0$, making it into a Lie bracket and turning $(\fg, [-,-]_0)$ into a Lie superalgebra, called a
\emph{contraction} of $(\fg, [-,-])$.  If $\phi_0 := \lim_{\varepsilon \to 0} \phi_\varepsilon$ exists and is invertible, the new Lie superalgebra is of
course isomorphic to the original one; hence one gets something new when $\phi_0$ either does not exist or fails to be invertible.

It is by no means guaranteed that the contraction of a minimally full Lie superalgebra will be minimally full, since contractions tend to increase the
kernel of the Lie bracket, thought of as a linear map.  Imposing that the contraction remain minimally full constrains the possible contractions.
Indeed, minimally full means that the Lie components $S^2\fg_1 \to \fg_0$ of both the original and contracted Lie brackets are isomorphisms.  This means
that the two brackets are related by an isomorphism of $\fg_0$.  Composing the family $\phi_\varepsilon$ with the inverse of this isomorphism, we may
assume that the bracket $S^2\fg_1 \to \fg_0$ remains unchanged under the contraction.  In other words, we can define $\phi_\varepsilon$ on $\fg_0$ in
terms of the restriction of $\phi_\varepsilon$ to $\fg_1$ as follows: for all $X,Y\in\fg_1$, we define $\phi_\varepsilon [X,Y] = [\phi_\varepsilon X,
\phi_\varepsilon Y]$, so that the component $S^2\fg_1 \to \fg_0$ of the Lie bracket remains unchanged as we change $\varepsilon$.  Since the contraction
is again minimally full, the above structure results say that the contracted Lie superalgebra is determined by some $\omega_0 \in
(\Lambda^2\fg_1^*)^{\fg_0}$.  A moment's thought reveals that $\omega_0 = \lim_{\varepsilon \to 0} \phi_\varepsilon^* \omega$, where $\omega$
characterises the original minimally full Lie superalgebra.

Let us illustrate this discussion.  Typically, the contraction maps $\phi_\varepsilon$ are simple rescalings $X \mapsto \varepsilon^{w(X)} X$, for some
weight function $w:\fg \to \RR$.  Let us decompose $\fg_1 = \fg_1^+ \oplus \fg_1^-$, which induces a decomposition $\fg_0 = \fg_0^{++} \oplus \fg_0^{+-}
\oplus \fg_0^{--}$, where the Lie bracket yields isomorphisms $S^2\fg_1^\pm \to \fg_0^{\pm\pm}$ and $\fg_1^+ \otimes \fg_1^- \to \fg_0^{+-}$.  Let us
now define the odd component $w: \fg_1 \to \RR$ of the weight function by
\begin{equation}
  w(X) =
  \begin{cases}
    1~, & X \in \fg_1^+\\
    0~, & X \in \fg_1^-~,
  \end{cases}
\end{equation}
which fixes the even component $w: \fg_0 \to \RR$ to be
\begin{equation}
  w(X) =
  \begin{cases}
    2~, & X \in \fg_0^{++}\\
    1~, & X \in \fg_0^{+-}\\
    0~, & X \in \fg_0^{--}~,
  \end{cases}
\end{equation}
for the contraction to be minimally full.  The contracted Lie brackets are of the form \eqref{eq:minfull} with $\omega$ the restriction of $\omega$ to
$\fg_1^-$.  In other words, if we decompose $\omega = \omega_{++} + \omega_{--} + \omega_{+-}$, then the contraction only retains $\omega_{--}$.  In
particular, the generators in $\fg_0^{++}$ become central.  This means that $\omega_0 = \omega_{--}$ now has rank $\dim \fg_1^-$, which can be any
\emph{even} number $0\leq r \leq \dim\fg_1$.

Returning briefly to the lack of existence of a maximal superalgebra for the maximally supersymmetric waves, we recall that it is known \cite{Limits}
that the plane-wave limit induces a contraction of the Killing superalgebra and it was shown explicitly in \cite{HatKamiSaka} that the Killing
superalgebras of the maximally supersymmetric Freund--Rubin and plane-wave backgrounds are related by a contraction.  It might therefore be again
surprising that whereas we have shown that the maximal extension of the Killing superalgebra of the maximally supersymmetric Freund--Rubin backgrounds
exists and is isomorphic to $\fosp(1|32)$, the Killing superalgebra of the maximally supersymmetric plane waves does not admit a minimally full
extension.  A resolution of this puzzle is to be found in the fact that for the contraction of a minimally full Lie superalgebra to remain minimally
full, the contraction map $\phi_\varepsilon$ on $\fm_0$ is determined by the one in $\fm_1$ (up to isomorphism).  A close analysis of the explicit
contraction in \cite{HatKamiSaka} shows that this contraction is not compatible with the minimally full condition: indeed, not every generator in
$\fk_0^{++}$ rescales the same way.  It is nevertheless a natural question to ask what are the possible contractions of $\fosp(1|32)$ which are
compatible with the plane-wave limit.

\bibliographystyle{utphys}
\bibliography{AdS,AdS3,ESYM,Sugra,Geometry,Algebra}

\end{document}